\documentclass[pdflatex,sn-mathphys,noend]{sn-jnl}

\usepackage{times}
\usepackage{url}
\usepackage{float}
\restylefloat{table}
\usepackage{hyperref}
\usepackage{bigstrut}
\usepackage{listings}
\usepackage{xcolor}
\usepackage{graphicx}
\usepackage{changepage}
\usepackage{algpseudocode}
\usepackage{algorithm}
\usepackage{float}
\usepackage{adjustbox}
\usepackage{bigstrut}
\usepackage{multirow}
\usepackage{color, colortbl}
\usepackage{comment}
\usepackage{algorithmicx}

\jyear{2023}%

\theoremstyle{thmstyleone}%
\theoremstyle{thmstyletwo}%

\theoremstyle{thmstylethree}%

\begin{document}

\title[Linear Index for Logarithmic Search-Time for any String  under any Internal Node in Suffix Trees]{Linear Index for Logarithmic Search-Time for any String  under any Internal Node in Suffix Trees}


\author*[1]{\fnm{Anas} \sur{Al-okaily}}\email{AA.12682@khcc.jo}



\affil*[1]{\orgdiv{Department of Cell Therapy and Applied Genomics}, \orgname{King Hussein Cancer Center}, \orgaddress{\city{Amman}, \country{Jordan}}}


\abstract{
Suffix trees are key and efficient data structure for solving string problems. A suffix tree is a compressed trie containing all the suffixes of a given text of length $n$ with a linear construction cost. In this work, we introduce an algorithm to build a linear index that allows finding a pattern of any length under any internal node in a suffix tree in $O(log_2 n)$ time.
}





\maketitle

\section{Introduction}\label{sec1}
Suffix trees $(STs)$ are widely used data structure for solving string problems especially problems related to pattern matching \cite{gusfield1997algorithms}. The cost for building suffix trees for a text ($T$) of length $n$ over an alphabet of size $\Sigma$ is linear (optimal). The tree can be used then to search for exact match for a given pattern in $O(l)$ (where $l$ is the length of the pattern) by walking the pattern in the $ST$ starting from the root node. However, many other problems including approximate pattern matching involve starting the search process from any internal node of suffix tree not only the root node. More reviews regarding this problem and its proposed solutions can be found in other studies \cite{Al-okaily2021.10.25.465764, kucherov2016approximate, hakak2019exact}. In this work and once the end node of walking a pattern $p$ starting from the root node of $ST$ is computed (with cost of $O(l)$ time), we propose a linear index that allows searching for $p$ under any internal node in $ST$ in $O(log_2 n)$ time. This index will contribute in resolving many string problems and mainly the approximate pattern matching problem. 

In our earlier algorithms \cite{Al-okaily2021.10.25.465764}, we introduced the concepts of base paths and base suffixes within the suffix tree structure \cite{ukkonen1995line}. Both algorithm are built at the top of suffix tree and another tree structure (derived from suffix trees) named as \textit{OSHR} tree. By using base suffix algorithm, indexing all strings under all internal nodes in $ST$ can be in linear space and time ($O(n)$). While, by using base paths algorithm the indexing all paths between each internal node and its descendant internal nodes can be also computed in linear time and space. In this work, we uses both algorithms with minor modification and with the use of additional node concepts. 

\section{Methods}\label{sec2}
At first, we will describe proposed index and how to build it. Then, the search process will be stated. 

\begin{figure}
    \centering
    
    \includegraphics[width=5cm]{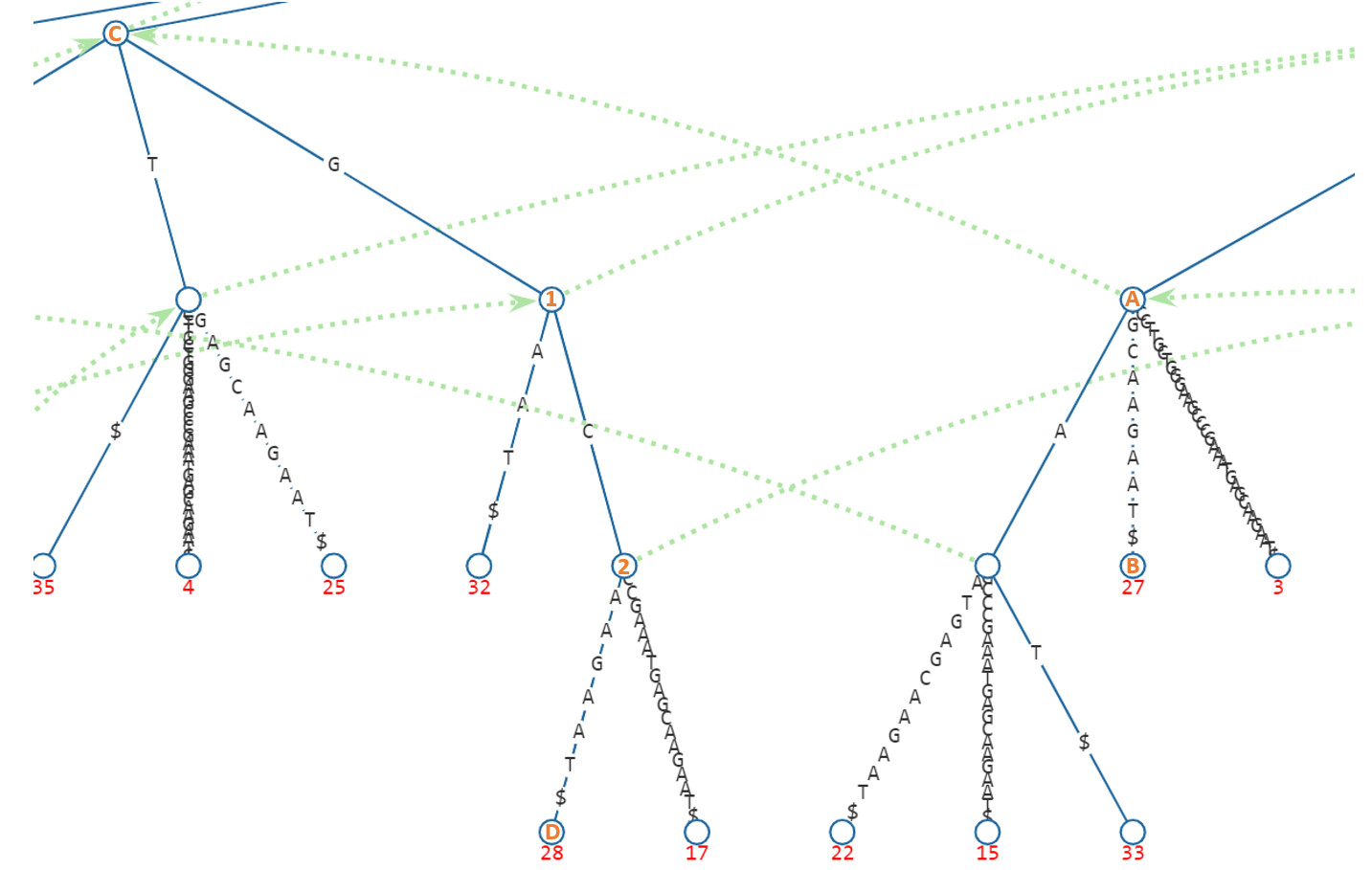}
    \includegraphics[width=5cm]{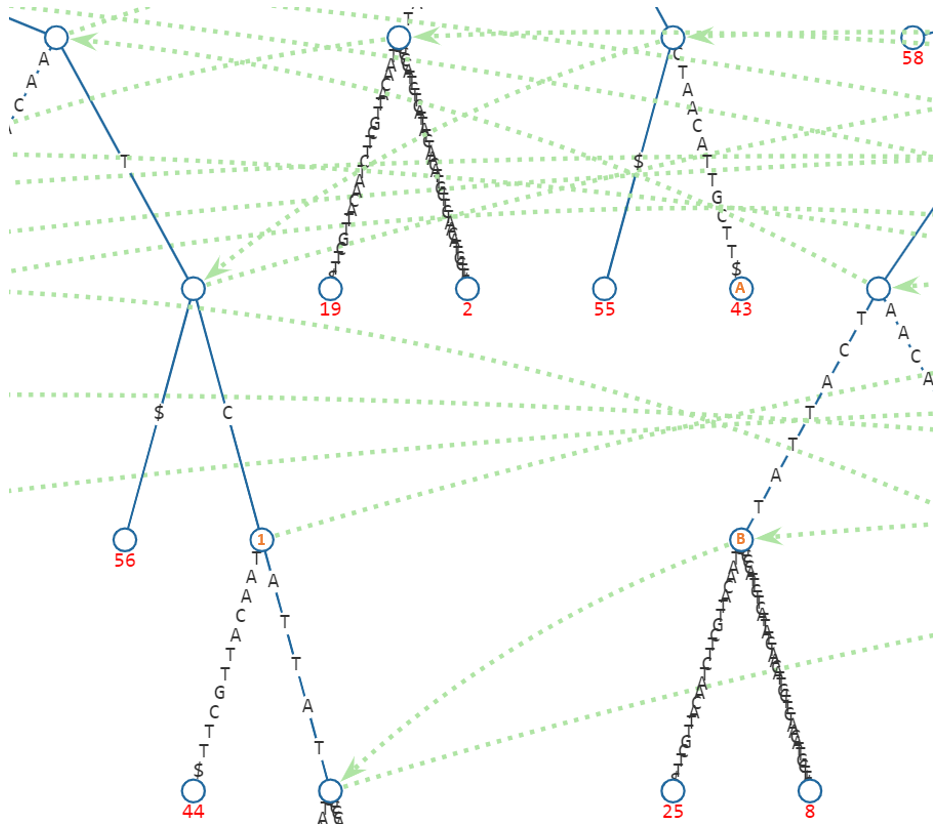}
    \caption{Node 1 in left picture is an example of Hanadi node; node A is a reference leaf node for node 1 (and hence is a reference node for Hanadi node); node 2 is not a Hanadi node as it's an \textit{OSHR} internal node and it must be already indexed by the index of base paths. Node 1 in right picture is an example of Srivastava node as it's an \textit{OSHR} leaf node, has reference leaf node (node A), and has reference internal node (node B); node A hence is also a reference node for Srivastava node.}
    \label{fig:enter-label}
\end{figure}

\subsection{Building the index}

The proposed index named as \textit{OT\_index} involves:
\begin{itemize}
    \item A minor modification of the computation of base suffixes as was computed in our earlier work \cite{Al-okaily2021.10.25.465764} as follows. The base suffixes (if any) at each internal node in $ST$ is derived from whether reference internal node or reference leaf node. Here we will compute and store at each internal node the base suffixes that are derived from reference leaf nodes only (the upper bound of the number of base suffixes derived from reference leaf node at any internal node is $\Sigma$ as was shown in \cite{Al-okaily2021.10.25.465764}).
    \item Three sub-indexes merged together: sub-index of base paths as was defined and described in \cite{Al-okaily2021.10.25.465764} (with minor modifications) and two sub-indexes derived from indexing two new types of nodes as will be described next.

\end{itemize}
In order to define the new two types of nodes, refer to the definitions of \textit{OSHR} internal node, \textit{OSHR} leaf node, reference internal nodes, and reference leaf nodes which are provided in \cite{Al-okaily2021.10.25.465764}.

\textbf{Definition 1}: Let A be a leaf node in \textit{ST} with suffix index $x$, B is the parent node of node A, C is the leaf node with suffix index $x+1$, and D is the parent node of node C. If the node suffix-linked from node B is not node D, then denote each node between node C and node D that is an \textit{OSHR} internal node as \textit{Hanadi} node and denote node A as a \textit {reference node for Hanadi node}. 

\textbf{Definition 2}: Let A be a leaf node in \textit{ST} with suffix index $x$, B is the parent node of node A, C is the leaf node with suffix index $x+1$, and D is the parent node of node C. If the node suffix-linked from node B is not node D, then denote each node between node C and node D that is an \textit{OSHR} leaf node and has at least one reference internal node as \textit{Srivastava} node and denote node A as a \textit {reference node for Srivastava node}. 

Figure 1 shows an example of Hanadi node, reference node for Hanadi node, Srivastava node, and reference node for Srivastava node. Clearly, Hanadi node can have more than one reference nodes and reference node for Hanadi nodes can have more than one Hanadi nodes. The same for Srivastava nodes and reference node for Srivastava nodes.

Algorithm 1 and Listing 2 in Supplementary Information show how \textit{OT\_index} can be built. 

\begin{algorithm*}
\tiny
\caption{Building the \textit{OT\_index} using the three sub-indexes}
\begin{algorithmic}[1]

\State Create and initialize a dictionary named as \textit{Indexed\_Paths}.
\State Build \textit{ST}.
\State Build \textit{OSHR} tree.
\State Traverse $ST$ and compute base suffixes derived from reference leaf nodes, find Hanadi nodes and reference nodes for Hanadi nodes (make sure to add each Hanadi node into a list at its reference node), find Srivastava nodes and reference nodes for Srivastava nodes (make sure to add each Srivastava node into a list at its reference node), and create a list at each internal node named as \textit{OT\_indexes} in which the \textit{OT} indexes using the three sub-indexes will be stored. 
\State Traverse $ST$ from left to right and create three lists: a list for storing left-to-right \textit{OSHR} leaf nodes, a list for storing left-to-right reference nodes for Hanadi nodes, and a list for storing left-to-right reference nodes for Srivastava nodes. This includes creating at each internal node left and right offsets correspondent to each list where left offset equals to the index of leftmost-leaf-node in the correspondent list and right offset is the index of rightmost leaf node in the correspondent list. These lists will be used during the indexing process and are needed to avoid non-linear computations that will be needed in finding (under each internal nodes) \textit{OSHR} leaf nodes, reference nodes for Hanadi nodes, and reference nodes for Srivastava nodes.

\Loop { through a postorder traversal of \textit{OSHR} tree and at each visited internal node $v$ (node $v$ is considered here as the top base node of the base path)}:

\For {each reference node for Hanadi/Srivastava nodes $e$ under node $v$} 
\For {each Hanadi/Srivastava stored at $e$}
\State Let $x$ = depth of parent node of Hanadi/Srivastava node - depth of node \textit{v.suffix-link}, and $y$ = depth of Hanadi/Srivastava node - depth of node \textit{v.suffix-link}. 
\State Find the last extent path of the path between node \textit{v.suffix-link} and Hanadi/Srivastava nodes (last extent path is the path in which the top base node is root node and its label is exact match of the base path).
\State Add an index consists of a tuple of (left\_OT\_index  of node $v$, right\_OT\_index  of node $v$, node $v$, and suffix index of reference nodes for Hanadi/Srivastava). 

\textit{OT\_indexes} in each node in last extent path with depth in range $x+1$ to $y$. 
\State Add the combination of \textit{v.suffix-link} and Hanadi/Srivastava nodes into \textit{Indexed\_Paths} dictionary (this dictionary will be used in next step). 
\EndFor
\EndFor

\For {each base path under node $v$, index base paths under node $v$ as described in \cite{Al-okaily2021.10.25.465764} with the following modification}
\If {the combination of top base node (node $v$) and bottom base node in \textit{Indexed\_Paths} dictionary} 
\State do nothing as this path was already indexed using the indexing of Hanadi nodes or Srivastava nodes. 
\Else 
\State Let's us denote the bottom base node of base path as node $a$ and the bottom base node of last extent path as node $b$. 
\State In the algorithm described in \cite{Al-okaily2021.10.25.465764}, we firstly find the last extent path of the base path (path between node $v$ and node $a$), then add an index into the list \textit{OT\_indexes} of node $b$. The index consists of a tuple of (left\_OT\_index and right\_OT\_index of node $v$, node $v$, suffix index of any leaf node under node $b$, node $b$, and a string indicates this is a base path index). While here and let $x$ = depth of parent node of node $b$ - depth of node $v$ and $y$ = depth of node $b$ - depth of node $v$, we will add the index into the list \textit{OT\_indexes} in each node in last extent path with depth in range $x+1$ to $y$. 
\EndIf
\If {node $b$ of base path has reference internal node $r$ and at least one of the nodes suffix-linked to node $v$ is an ancestor to node $r$}
\State Exclude this base path (between node $v$ and node $b$) from the indexing process as this base paths must have been already covered. 
\EndIf \EndFor \EndLoop 

\State Traverse \textit{ST} or \textit{OSHR} tree and sort each list \textit{OT\_indexes} at each internal node based on the first two elements (left\_OT\_index and right\_OT\_index values). 

\Comment{If the length of the inputted patterns is fixed (let's say the length is $l$), then the size of indexing process can be highly reduced as follows. First, consider only base paths or paths between node \textit{v.suffix-link} and Hanadi/Srivastava nodes \textbf{where} $x+1 \lt l \leq y$. For those paths, add the index into the list \textit{OT\_indexes} of node with depth equals to $l$ in the last extent path of those paths. If last extent path does not have a node with depth equal to $l$, then add the index into the list \textit{OT\_indexes} of the \textit{first} node with depth greater than $l$. This also can be applied accordingly when the length of the inputted patterns is $l$ or less or is within a range of two values.}

\end{algorithmic}
\end{algorithm*}

\subsection{Search process}

The algorithm on how to compute the search using \textit{OT\_index} is stated in Algorithm 2. 
\begin{algorithm*}
\tiny
\caption{Searching algorithm for a pattern under any internal node}
\begin{algorithmic}[1]
\State Let the end node of walking pattern $p$ as exact match starting from the root node be node $e$.
\State Let the internal node that the search process will start from is node $i$.
\If {node $e$ is a leaf node}
\State Finding if $p$ presented under node $i$ as exact match can be naively computed in constant time (description is provided in Supplementary Information). Hence, no need to use the \textit{OT\_index} or any index.
\Else 
\If {the list \textit{OT\_indexes} at node $e$ is empty i.e. the path between root node and node $e$ is a base path}. 
\State Check if any base suffix in the list of base suffixes derived from reference leaf nodes stored at node $e$ occurred under node $i$. This check can be computed in $O(\Sigma)$ time trivially (by iterating over each $O(\Sigma)$ base suffix) and in $O(log_2\Sigma)$ time using non-trivial algorithm named as Mandoiu algorithm (as provided in Listing 3 in Supplementary Information).
\Else   
    \State Perform binary search for the value of \textit{right\_OT\_index} of node $i$ in the list \textit{OT\_indexes} at node $e$. If binary search found an index, use this index to find the matching as stated in Listing 4 in Supplementary Information. 
\EndIf
\EndIf

\end{algorithmic}
\end{algorithm*}

\section{Results}\label{sec3}

\textit{OT\_index} was implemented using Python3 language. Eight genomes were used (ranging in size from 500KB to 100MB) in order to show the costs for the \textit{OT\_index} over different genome-sizes and to compare the search process using \textit{OT\_index} and search process using walk algorithm for different patterns and starting from different internal nodes. The genomes are: WS1 bacterium JGI 0000059-K21 (Bacteria), Astrammina rara (Protist), Nosema ceranae (Fungus), Cryptosporidium parvumIowa II (Protist), Spironucleus salmonicida (Protist), Tieghemostelium lacteum (Protist), Fusarium graminearumPH-1 (Fungus), Salpingoeca rosetta (Protist), and Chondrus crispus (Algae). As a preprocessing step, header lines and newlines were removed from the fasta file and any small letter nucleotide were capitalized. This generates a one line genome file for each fasta file where all nucleotide are capital. The python script used for this preprocessing step is provided at \url{https://github.com/aalokaily/Searching_using_OT_index/}. 

\begin{table}[htbp]
  \centering
  \caption{The size of each sub-index of the three sub-indexes (base paths, Hanadi nodes, and Srivastava nodes) for each genome.}
  \begin{adjustbox}{max width=\textwidth}

\begin{tabular}{|l|r|r|r|r|r|r|r|r|r|}
\hline
\rowcolor[rgb]{ .851,  .851,  .851} \multicolumn{1}{|c|}{\textbf{Genome}} &
  \multicolumn{1}{c|}{\textbf{WS1 bacterium}} &
  \multicolumn{1}{c|}{\textbf{Astrammina}} &
  \multicolumn{1}{c|}{\textbf{Nosema}} &
  \multicolumn{1}{c|}{\textbf{Cryptosporidium}} &
  \multicolumn{1}{c|}{\textbf{Spironucleus}} &
  \multicolumn{1}{c|}{\textbf{Tieghemostelium}} &
  \multicolumn{1}{c|}{\textbf{Fusarium}} &
  \multicolumn{1}{c|}{\textbf{Salpingoeca}} &
  \multicolumn{1}{c|}{\textbf{Chondrus}}
  \bigstrut\\
\hline
\rowcolor[rgb]{ .851,  .851,  .851} GenBank Accession &
  \cellcolor[rgb]{ 1,  1,  1} GCA000398605.1 &
  \cellcolor[rgb]{ 1,  1,  1} GCA000211355.2 &
  \cellcolor[rgb]{ 1,  1,  1} GCA000988165.1 &
  \cellcolor[rgb]{ 1,  1,  1} GCA000165345.1 &
  \cellcolor[rgb]{ 1,  1,  1} GCA000497125.1 &
  \cellcolor[rgb]{ 1,  1,  1} GCA001606155.1 &
  \cellcolor[rgb]{ 1,  1,  1} GCF000240135.3 &
  \cellcolor[rgb]{ 1,  1,  1} GCA000188695.1 &
  \cellcolor[rgb]{ 1,  1,  1} GCA000350225.2
  \bigstrut\\
\hline
\rowcolor[rgb]{ .851,  .851,  .851} No. of alphabets  &
  \cellcolor[rgb]{ 1,  1,  1} 5 &
  \cellcolor[rgb]{ 1,  1,  1} 5 &
  \cellcolor[rgb]{ 1,  1,  1} 4 &
  \cellcolor[rgb]{ 1,  1,  1} 15 &
  \cellcolor[rgb]{ 1,  1,  1} 5 &
  \cellcolor[rgb]{ 1,  1,  1} 4 &
  \cellcolor[rgb]{ 1,  1,  1} 5 &
  \cellcolor[rgb]{ 1,  1,  1} 5 &
  \cellcolor[rgb]{ 1,  1,  1} 5
  \bigstrut\\
\hline
\rowcolor[rgb]{ .851,  .851,  .851} No. of nuc/leaf nodes &
  \cellcolor[rgb]{ 1,  1,  1} 509,552 &
  \cellcolor[rgb]{ 1,  1,  1} 1,450,096 &
  \cellcolor[rgb]{ 1,  1,  1} 5,690,749 &
  \cellcolor[rgb]{ 1,  1,  1} 9,102,325 &
  \cellcolor[rgb]{ 1,  1,  1} 12,954,589 &
  \cellcolor[rgb]{ 1,  1,  1} 23,375,663 &
  \cellcolor[rgb]{ 1,  1,  1} 36,458,047 &
  \cellcolor[rgb]{ 1,  1,  1} 55,440,310 &
  \cellcolor[rgb]{ 1,  1,  1} 104,980,421
  \bigstrut\\
\hline
\rowcolor[rgb]{ .851,  .851,  .851} No. of Internal nodes &
  \cellcolor[rgb]{ 1,  1,  1} 328,917 &
  \cellcolor[rgb]{ 1,  1,  1} 926,087 &
  \cellcolor[rgb]{ 1,  1,  1} 3,849,880 &
  \cellcolor[rgb]{ 1,  1,  1} 5,931,181 &
  \cellcolor[rgb]{ 1,  1,  1} 8,684,514 &
  \cellcolor[rgb]{ 1,  1,  1} 15,606,221 &
  \cellcolor[rgb]{ 1,  1,  1} 22,972,065 &
  \cellcolor[rgb]{ 1,  1,  1} 37,480,799 &
  \cellcolor[rgb]{ 1,  1,  1} 81,909,252
  \bigstrut\\
\hline
\rowcolor[rgb]{ .851,  .851,  .851} Size of index of base paths &
  \cellcolor[rgb]{ 1,  1,  1} 1,470,339 &
  \cellcolor[rgb]{ 1,  1,  1} 4,393,219 &
  \cellcolor[rgb]{ 1,  1,  1} 20,757,070 &
  \cellcolor[rgb]{ 1,  1,  1} 33,266,692 &
  \cellcolor[rgb]{ 1,  1,  1} 47,876,615 &
  \cellcolor[rgb]{ 1,  1,  1} 92,141,518 &
  \cellcolor[rgb]{ 1,  1,  1} 137,440,414 &
  \cellcolor[rgb]{ 1,  1,  1} 251,548,596 &
  \cellcolor[rgb]{ 1,  1,  1} 339,798,161
  \bigstrut\\
\hline
\rowcolor[rgb]{ .851,  .851,  .851} Size of index of Hanadi nodes  &
  \cellcolor[rgb]{ 1,  1,  1} 1,212,706 &
  \cellcolor[rgb]{ 1,  1,  1} 3,475,995 &
  \cellcolor[rgb]{ 1,  1,  1} 16,959,305 &
  \cellcolor[rgb]{ 1,  1,  1} 28,231,517 &
  \cellcolor[rgb]{ 1,  1,  1} 35,524,630 &
  \cellcolor[rgb]{ 1,  1,  1} 76,007,349 &
  \cellcolor[rgb]{ 1,  1,  1} 111,528,426 &
  \cellcolor[rgb]{ 1,  1,  1} 218,306,328 &
  \cellcolor[rgb]{ 1,  1,  1} 232,633,080
  \bigstrut\\
\hline
\rowcolor[rgb]{ .851,  .851,  .851} Size of index of Srivastava nodes  &
  \cellcolor[rgb]{ 1,  1,  1} 219,229 &
  \cellcolor[rgb]{ 1,  1,  1} 766,251 &
  \cellcolor[rgb]{ 1,  1,  1} 4,539,219 &
  \cellcolor[rgb]{ 1,  1,  1} 5,086,705 &
  \cellcolor[rgb]{ 1,  1,  1} 11,579,129 &
  \cellcolor[rgb]{ 1,  1,  1} 17,842,779 &
  \cellcolor[rgb]{ 1,  1,  1} 18,520,263 &
  \cellcolor[rgb]{ 1,  1,  1} 51,221,684 &
  \cellcolor[rgb]{ 1,  1,  1} 113,869,263
  \bigstrut\\
\hline
\rowcolor[rgb]{ .851,  .851,  .851} \textbf{Total size of three indexes} &
  \cellcolor[rgb]{ 1,  1,  1} \textbf{2,902,274} &
  \cellcolor[rgb]{ 1,  1,  1} \textbf{8,635,465} &
  \cellcolor[rgb]{ 1,  1,  1} \textbf{42,255,594} &
  \cellcolor[rgb]{ 1,  1,  1} \textbf{66,584,914} &
  \cellcolor[rgb]{ 1,  1,  1} \textbf{94,980,374} &
  \cellcolor[rgb]{ 1,  1,  1} \textbf{185,991,646} &
  \cellcolor[rgb]{ 1,  1,  1} \textbf{267,489,103} &
  \cellcolor[rgb]{ 1,  1,  1} \textbf{521,076,608} &
  \cellcolor[rgb]{ 1,  1,  1} \textbf{686,300,504}
  \bigstrut\\
\hline
\end{tabular}%

  \end{adjustbox}
  \label{tab:addlabel}%
\end{table}%

Table 1 shows the metadata of the suffix trees built for each genome and the size of each sub-index. Note that the size of each sub-index always is less than $O(\Sigma n)$. In addition, the order of sizes of the three sub-indexes is always: sub-index using base paths, sub-index using Hanadi nodes, then sub-index using Srivastava nodes. This leads to a conclusion that the upper bound of the size of sub-index using base paths and using Hanadi nodes is $O(\Sigma n)$ and $O(n)$ for sub-index using Srivastava nodes. Hence, the upper bound for \textit{OT\_index} in total is $O(\Sigma n)$. As a result, the binary search in \textit{OT\_index} will cost $log_2n$ time (given that OT\_indexes are distributive over almost all internal nodes, hence the expected size of \textit{OT\_index} list at each of most internal nodes is much less than $n$, therefore the binary search will be faster than $O(log_2n)$) (in fact, the deeper the internal node is the less size its \textit{OT\_index} list).

\begin{table}[htbp]
  \centering
  \caption{The result of testing naive algorithm (search by walking) and the proposed algorithm (three sub-indexes). The complexity is defined here as the total number of internal nodes under the starting nodes.}
  \begin{adjustbox}{max width=\textwidth}
    \begin{tabular}{|r|r|r|r|r|r|r|}
    \hline
    \rowcolor[rgb]{ .651,  .651,  .651} \multicolumn{1}{|p{4.215em}|}{\textbf{Depth}} &
      \multicolumn{1}{p{7.785em}|}{\textbf{No of nodes at depth}} &
      \multicolumn{1}{p{5.43em}|}{\textbf{No of collected patterns}} &
      \multicolumn{1}{p{7.285em}|}{\textbf{No of starting nodes at depth}} &
      \multicolumn{1}{p{6.07em}|}{\textbf{Complexity}} &
      \multicolumn{1}{p{6.785em}|}{\textbf{Search time (sec) using \textit{OT\_index}}} &
      \multicolumn{1}{p{7.145em}|}{\textbf{Search time (sec) using walk algorithm}}
      \bigstrut\\
    \hline
    \rowcolor[rgb]{ .651,  .651,  .651} \textbf{1} &
      \cellcolor[rgb]{ 1,  1,  1}                               53  &
      \cellcolor[rgb]{ 1,  1,  1}              8,126  &
      \cellcolor[rgb]{ 1,  1,  1}                             53  &
      \cellcolor[rgb]{ 1,  1,  1}                     338  &
      \cellcolor[rgb]{ 1,  1,  1} 0.57499 &
      \cellcolor[rgb]{ 1,  1,  1} \textbf{0.45979}
      \bigstrut\\
    \hline
    \rowcolor[rgb]{ .651,  .651,  .651} \textbf{2} &
      \cellcolor[rgb]{ 1,  1,  1}                             327  &
      \cellcolor[rgb]{ 1,  1,  1}              8,126  &
      \cellcolor[rgb]{ 1,  1,  1}                           327  &
      \cellcolor[rgb]{ 1,  1,  1}                 1,659  &
      \cellcolor[rgb]{ 1,  1,  1} \textbf{1.11445} &
      \cellcolor[rgb]{ 1,  1,  1} 1.80536
      \bigstrut\\
    \hline
    \rowcolor[rgb]{ .651,  .651,  .651} \textbf{3} &
      \cellcolor[rgb]{ 1,  1,  1}                         1,570  &
      \cellcolor[rgb]{ 1,  1,  1}              8,126  &
      \cellcolor[rgb]{ 1,  1,  1}                       1,570  &
      \cellcolor[rgb]{ 1,  1,  1}                 4,948  &
      \cellcolor[rgb]{ 1,  1,  1} \textbf{4.14428} &
      \cellcolor[rgb]{ 1,  1,  1} 7.15639
      \bigstrut\\
    \hline
    \rowcolor[rgb]{ .651,  .651,  .651} \textbf{4} &
      \cellcolor[rgb]{ 1,  1,  1}                         4,568  &
      \cellcolor[rgb]{ 1,  1,  1}              8,126  &
      \cellcolor[rgb]{ 1,  1,  1}                       3,759  &
      \cellcolor[rgb]{ 1,  1,  1}               11,817  &
      \cellcolor[rgb]{ 1,  1,  1} \textbf{10.09327} &
      \cellcolor[rgb]{ 1,  1,  1} 21.05655
      \bigstrut\\
    \hline
    \rowcolor[rgb]{ .651,  .651,  .651} \textbf{5} &
      \cellcolor[rgb]{ 1,  1,  1}                       12,201  &
      \cellcolor[rgb]{ 1,  1,  1}              8,126  &
      \cellcolor[rgb]{ 1,  1,  1}                       9,000  &
      \cellcolor[rgb]{ 1,  1,  1}               30,887  &
      \cellcolor[rgb]{ 1,  1,  1} \textbf{24.07914} &
      \cellcolor[rgb]{ 1,  1,  1} 55.65963
      \bigstrut\\
    \hline
    \rowcolor[rgb]{ .651,  .651,  .651} \textbf{6} &
      \cellcolor[rgb]{ 1,  1,  1}                       40,166  &
      \cellcolor[rgb]{ 1,  1,  1}              8,126  &
      \cellcolor[rgb]{ 1,  1,  1}                       9,000  &
      \cellcolor[rgb]{ 1,  1,  1}               34,494  &
      \cellcolor[rgb]{ 1,  1,  1} \textbf{24.04910} &
      \cellcolor[rgb]{ 1,  1,  1} 57.21128
      \bigstrut\\
    \hline
    \rowcolor[rgb]{ .651,  .651,  .651} \textbf{7} &
      \cellcolor[rgb]{ 1,  1,  1}                     149,447  &
      \cellcolor[rgb]{ 1,  1,  1}              8,126  &
      \cellcolor[rgb]{ 1,  1,  1}                       9,000  &
      \cellcolor[rgb]{ 1,  1,  1}               34,396  &
      \cellcolor[rgb]{ 1,  1,  1} \textbf{23.57629} &
      \cellcolor[rgb]{ 1,  1,  1} 47.46837
      \bigstrut\\
    \hline
    \rowcolor[rgb]{ .651,  .651,  .651} \textbf{8} &
      \cellcolor[rgb]{ 1,  1,  1}                     567,245  &
      \cellcolor[rgb]{ 1,  1,  1}              8,126  &
      \cellcolor[rgb]{ 1,  1,  1}                       9,000  &
      \cellcolor[rgb]{ 1,  1,  1}               33,162  &
      \cellcolor[rgb]{ 1,  1,  1} \textbf{23.40210} &
      \cellcolor[rgb]{ 1,  1,  1} 38.98778
      \bigstrut\\
    \hline
    \rowcolor[rgb]{ .651,  .651,  .651} \textbf{9} &
      \cellcolor[rgb]{ 1,  1,  1}                 2,005,293  &
      \cellcolor[rgb]{ 1,  1,  1}              8,126  &
      \cellcolor[rgb]{ 1,  1,  1}                       9,000  &
      \cellcolor[rgb]{ 1,  1,  1}               30,635  &
      \cellcolor[rgb]{ 1,  1,  1} \textbf{23.52925} &
      \cellcolor[rgb]{ 1,  1,  1} 31.17348
      \bigstrut\\
    \hline
    \rowcolor[rgb]{ .651,  .651,  .651} \textbf{10} &
      \cellcolor[rgb]{ 1,  1,  1}                 6,110,918  &
      \cellcolor[rgb]{ 1,  1,  1}              8,126  &
      \cellcolor[rgb]{ 1,  1,  1}                       9,000  &
      \cellcolor[rgb]{ 1,  1,  1}               26,698  &
      \cellcolor[rgb]{ 1,  1,  1} \textbf{23.71234} &
      \cellcolor[rgb]{ 1,  1,  1} 24.80497
      \bigstrut\\
    \hline
    \rowcolor[rgb]{ .651,  .651,  .651} \textbf{11} &
      \cellcolor[rgb]{ 1,  1,  1}               16,414,131  &
      \cellcolor[rgb]{ 1,  1,  1}              8,126  &
      \cellcolor[rgb]{ 1,  1,  1}                       9,000  &
      \cellcolor[rgb]{ 1,  1,  1}               20,563  &
      \cellcolor[rgb]{ 1,  1,  1} 24.02936 &
      \cellcolor[rgb]{ 1,  1,  1} \textbf{20.63834}
      \bigstrut\\
    \hline
    \rowcolor[rgb]{ .651,  .651,  .651} \textbf{12} &
      \cellcolor[rgb]{ 1,  1,  1}               31,975,091  &
      \cellcolor[rgb]{ 1,  1,  1}              8,126  &
      \cellcolor[rgb]{ 1,  1,  1}                       9,000  &
      \cellcolor[rgb]{ 1,  1,  1}               13,981  &
      \cellcolor[rgb]{ 1,  1,  1} 24.32627 &
      \cellcolor[rgb]{ 1,  1,  1} \textbf{18.43981}
      \bigstrut\\
    \hline
    \rowcolor[rgb]{ .651,  .651,  .651} \textbf{13} &
      \cellcolor[rgb]{ 1,  1,  1}               30,817,995  &
      \cellcolor[rgb]{ 1,  1,  1}              8,126  &
      \cellcolor[rgb]{ 1,  1,  1}                       9,000  &
      \cellcolor[rgb]{ 1,  1,  1}                 9,275  &
      \cellcolor[rgb]{ 1,  1,  1} 24.51689 &
      \cellcolor[rgb]{ 1,  1,  1} \textbf{17.25809}
      \bigstrut\\
    \hline
    \rowcolor[rgb]{ .651,  .651,  .651} \textbf{14} &
      \cellcolor[rgb]{ 1,  1,  1}               17,769,354  &
      \cellcolor[rgb]{ 1,  1,  1}              8,126  &
      \cellcolor[rgb]{ 1,  1,  1}                       9,000  &
      \cellcolor[rgb]{ 1,  1,  1}                 6,804  &
      \cellcolor[rgb]{ 1,  1,  1} 24.56814 &
      \cellcolor[rgb]{ 1,  1,  1} \textbf{16.68087}
      \bigstrut\\
    \hline
    \rowcolor[rgb]{ .651,  .651,  .651} \textbf{15} &
      \cellcolor[rgb]{ 1,  1,  1}                 8,556,089  &
      \cellcolor[rgb]{ 1,  1,  1}              8,126  &
      \cellcolor[rgb]{ 1,  1,  1}                       8,787  &
      \cellcolor[rgb]{ 1,  1,  1}                 5,390  &
      \cellcolor[rgb]{ 1,  1,  1} 24.23818 &
      \cellcolor[rgb]{ 1,  1,  1} \textbf{16.21155}
      \bigstrut\\
    \hline
    \rowcolor[rgb]{ .651,  .651,  .651} \textbf{16} &
      \cellcolor[rgb]{ 1,  1,  1}                 4,271,921  &
      \cellcolor[rgb]{ 1,  1,  1}              8,126  &
      \cellcolor[rgb]{ 1,  1,  1}                       8,314  &
      \cellcolor[rgb]{ 1,  1,  1}                 4,750  &
      \cellcolor[rgb]{ 1,  1,  1} 23.72229 &
      \cellcolor[rgb]{ 1,  1,  1} \textbf{15.73226}
      \bigstrut\\
    \hline
    \rowcolor[rgb]{ .651,  .651,  .651} \textbf{17} &
      \cellcolor[rgb]{ 1,  1,  1}                 2,502,348  &
      \cellcolor[rgb]{ 1,  1,  1}              8,126  &
      \cellcolor[rgb]{ 1,  1,  1}                       8,152  &
      \cellcolor[rgb]{ 1,  1,  1}                 4,301  &
      \cellcolor[rgb]{ 1,  1,  1} 23.45970 &
      \cellcolor[rgb]{ 1,  1,  1} \textbf{15.45235}
      \bigstrut\\
    \hline
    \rowcolor[rgb]{ .651,  .651,  .651} \textbf{18} &
      \cellcolor[rgb]{ 1,  1,  1}                 1,762,680  &
      \cellcolor[rgb]{ 1,  1,  1}              8,126  &
      \cellcolor[rgb]{ 1,  1,  1}                       8,102  &
      \cellcolor[rgb]{ 1,  1,  1}                 3,986  &
      \cellcolor[rgb]{ 1,  1,  1} 23.40955 &
      \cellcolor[rgb]{ 1,  1,  1} \textbf{15.33595}
      \bigstrut\\
    \hline
    \rowcolor[rgb]{ .651,  .651,  .651} \textbf{19} &
      \cellcolor[rgb]{ 1,  1,  1}                 1,433,875  &
      \cellcolor[rgb]{ 1,  1,  1}              8,126  &
      \cellcolor[rgb]{ 1,  1,  1}                       8,079  &
      \cellcolor[rgb]{ 1,  1,  1}                 3,903  &
      \cellcolor[rgb]{ 1,  1,  1} 23.30871 &
      \cellcolor[rgb]{ 1,  1,  1} \textbf{15.29407}
      \bigstrut\\
    \hline
    \rowcolor[rgb]{ .651,  .651,  .651} \textbf{20} &
      \cellcolor[rgb]{ 1,  1,  1}                 1,264,914  &
      \cellcolor[rgb]{ 1,  1,  1}              8,126  &
      \cellcolor[rgb]{ 1,  1,  1}                       8,071  &
      \cellcolor[rgb]{ 1,  1,  1}                 4,180  &
      \cellcolor[rgb]{ 1,  1,  1} 23.17237 &
      \cellcolor[rgb]{ 1,  1,  1} \textbf{15.23059}
      \bigstrut\\
    \hline
    \end{tabular}%
    \end{adjustbox}
  \label{tab:addlabel}%
\end{table}%

In order to benchmark the search process using \textit{OT\_index} and to compare it with the search process using walk algorithm starting from different internal nodes in $ST$, we conducted the following test. Starting by collecting 100 patterns for each length of 7, 10, 12, 15, 20, 25, 30, 35, 40, and 50 (total of 1,000 patterns) from each genome. These 100 patterns were extracted from the genome from position $10 * i * length$ where $i$ the number of pattern. As an example, for patterns of length 12, the first pattern must be extracted from the genome at position 120, the second at position 240, the third at position 360, and so on. If the end node of walking the pattern from root node is a leaf node, then disregard this pattern as the search for such a pattern starting from any internal node in $ST$ can be computed in a constant time and there will be no need to use \textit{OT\_index}. Note that the total number of extracted patterns for short genomes can be less than 100 (as indicated in Table 2). Next, each pattern of the 1,000 patterns collected from each genome was searched starting from up to 1,000 internal node at each depth of 1, 2, 3, ..., 20 in the $ST$ built for the genome. The outcomes of this test collectively for the nine genomes is shown in Table 2. 

Clearly, the search process using \textit{OT\_index} for depths 2-10 is much faster (almost half) than the search process using walk algorithm. This is justified as the complexity under internal nodes at these depths are the highest, hence the walking algorithm will be costly when compared to nodes at deeper depths and when compared to binary search with max cost of $O(log_2n)$ (given that again the size of \textit{OT\_index} list at any of most internal nodes is much less than $n$ hence the cost of the binary search is much less than $O(log_2n)$). In addition, given that the search process using \textit{OT\_index} is not beneficial at depths other than depths 2-10, the total number of internal nodes in depths 2-10 is still very high (these nodes are expected to be starting nodes for a search process) and they already are the nodes with highest complexity. Moreover, the search time using \textit{OT\_index} over depth 5 or more is slightly constant (while this is not the case for the search time using walk algorithm). This is due to the slight change that is caused by the $O(log_2n)$ cost for binary search. Finally, these conclusions are the same for each genome independently (tables for each genome are provided in Supplementary Information).

\backmatter

\bmhead{Supplementary information} There is Supplementary information. 

\bmhead{Acknowledgments} 
The naming of Hanadi nodes is to tribute Hanadi Al Tbeishat. The naming of Srivastava nodes to tribute Pramod Srivastsva (Department of Immunology, University of Connecticut). The naming of Mandoiu search method is to tribute Ion Mandoiu (Department of Computer Science, University of Connecticut). 

\bmhead{Code availability}
The algorithms proposed in this paper are implemented using Python3 programming language and available at 
\url{https://github.com/aalokaily/Searching_using_OT_index/}

\bibliography{sn-bibliography}

\end{document}